\documentstyle[prb,aps,epsf]{revtex}

\begin{document}
\title{An exact renormalization group  approach to frustrated magnets}
\author{M. Tissier,  B. Delamotte and D. Mouhanna}

\address{Laboratoire de Physique Th\'eorique et Hautes Energies. Universit\'es Paris
 VI-Pierre et Marie Curie -- Paris VII-Denis Diderot, 2 Place Jussieu, 75252 Paris Cedex
 05, France}

\maketitle

\begin{abstract}
Frustrated magnets are a notorious example where  usual
 perturbative methods fail.  Having recourse to an exact renormalization group approach,
 one gets a coherent picture of the physics of Heisenberg frustrated magnets everywhere
 between $d=2$ and $d=4$: all known perturbative results are recovered in a single
 framework, their apparent conflict is explained while the description of the phase
 transition in $d=3$ is found to be in good agreement with the experimental context.
\end{abstract}

\section{Frustrated Magnets}

One of the great challenges of present-day condensed matter physics is to 
understand the effects of competing interactions -- frustration -- in
three-dimensional spin systems. However, after twenty five years of
investigations, the question of the nature of the universality class for the phase
transition of the simplest frustrated model, the classical antiferromagnetic
Heisenberg model on a triangular lattice (AFHT model), is still strongly debated~\cite{kawamura10}. The main
feature of this system is that, due to frustration, its ground state  is given by a
configuration in which  the spins have a {\it non-collinear} order: on
an elementary triangular plaquette, they point at 120$^{\circ}$ one from another. A
consequence of this non-collinear order  is that, contrary to the case of
non-frustrated spin systems, the microscopic order parameter is no longer a single vector but a matrix, built from two {\it orthonormal} vectors
$(\vec{\phi}_1,\vec{\phi}_2)$,  playing a role similar to the staggered magnetization in non-frustrated antiferromagnetic systems. This implies a symmetry breaking scheme in which the group of spatial rotation is {\it fully} broken. In the case of the AFHT model it is given by $O(3)\otimes O(2)\to O(2)$~\cite{aza4}. This situation  raises the question of the nature of the phase transition for frustrated magnets in
three dimensions.  Experiments performed on materials supposed to belong to the AFHT
 universality class exhibit a critical behavior with exponents differing from those of the standard $O(N)$ universality class~\cite{collins}.  However the experimental situation calls for several important remarks. First, the critical exponents differ significantly from one material to another and also with those obtained by Monte Carlo simulations performed either directly on the AFHT model or on related models (see ref.[4] for a review). Second, for a given material, the scaling relations
 are violated at least by two standard deviations. These results are manifestly not
compatible with the existence of a true universality.

\section{Perturbative Field Theoretical Context}

From the theoretical point of view, this state of affairs raises the question
 of the ability of the field theoretical formalism to explain the
experimental and numerical  situation.  As in the $O(N)$ case, mainly two kinds of approaches have been used to tackle with this problem. We now display the results of
these studies.

\subsection{The Non-Linear Sigma Model Approach}

The first approach consists in a low-temperature expansion
performed around two dimensions on the suitable field theory, 
the $O(3)\otimes O(2)/ O(2)$ Non-Linear Sigma (NL$\sigma$) model~\cite{aza4} which is obtained by expressing the interaction between spins in terms of the Goldstone modes, the excitations associated to the transverse fluctuations of the magnetization. This is realized, as usual, by using the geometrical constraints of the ground state -- here the
orthonormality of $\vec{\phi}_1$ and $\vec{\phi}_2$ -- to get rid of the irrelevant -- longitudinal -- degrees of freedom. Formally, the resulting partition function writes: 
\begin{equation}
Z=\int D{\pi} \
\exp \left[ -{{1\over 2}\int d^dx\ g_{ij}(\pi)\
\partial_{\mu}\pi^i\partial_{\mu}\pi^j} \right]
\label{nonlinear}
\end{equation}
where the $\pi^i$'s represent the -- three -- Goldstone modes of the model,
$g_{ij}(\pi)$ encoding their interaction. The expression (\ref{nonlinear}) is
then used to evaluate the renormalization of the interaction between  the Goldstone modes in a double expansion in the temperature $T$ and in $\epsilon=d-2$.  In
the case of frustrated magnets, the result of this study is that
there exists, above two dimensions, a stable fixed point~\cite{aza4}. At this fixed
point, the symmetry breaking scheme is {\it enlarged} from
$O(3)\otimes O(2)/O(2)$ to
$O(3)\otimes O(3)/O(3)$. An important point is that the
renormalization group (RG) properties of a NL$\sigma$ model depend only
on the geometry of the order parameter space -- the coset $G/H$ -- viewed as a manifold equipped
with the metric $g_{ij}(\pi)$~\cite{friedan}. In particular, the $\beta$ functions for the
coupling constants entering in the model are given in terms of
geometrical quantities on $G/H$ like Ricci  -- $R_{ij}$ -- and Riemann --
$R_{i}^{  pqr}$ -- tensors on the manifold~\cite{friedan}: 
\begin{equation}
\beta _{ij}(g)= { \partial g_{ij}\over \partial l} =\epsilon g_{ij} -{1\over 2\pi} R_{ij} - {1\over8\pi^2}
R_{i}^{  pqr}R_{jpqr}\ .
\label{betafunction}
\end{equation}
This implies that the RG properties depend only on the $\it local$
properties of $G/H$: the Lie algebras of the  symmetry groups of the
high and low temperature phases. As a consequence, they are insensitive to its {\it global}
structure -- topology -- and to the  representation used to realized the
symmetry breaking scheme. It is known that the manifold $O(3)\otimes O(3)/O(3)$ has the same local structure as $O(4)/O(3)$ on which is built the NL$\sigma$ model relevant to the four-component {\it non-frustrated} vector model. As seen from the
NL$\sigma$ model calculations, the two above theories are thus equivalent
to all orders. Accordingly, the critical properties of the the AFHT model between two and four dimensions should be governed by the standard $O(4)/O(3)$ universality class~\cite{aza4}. This
is however in strong disagreement with both experimental and
numerical results as well as the weak-coupling analysis of the Landau-Ginzburg-Wilson (LGW) model of frustrated magnets around four dimensions.

\subsection{LGW model approach}

An alternative approach to that of the NL$\sigma$ model is to 
consider a suitable LGW representation of the physics of frustrated
magnets. This can be obtained, for instance,  by relaxing, via a potential, the
constraints defining the ground state:
\begin{equation}
Z=\int  D\vec{\phi}_1  D\vec{\phi}_2\  e^{-\int d^Dx
{1\over2}\left(\nabla
\vec{\phi_1}^2 + \nabla \vec{\phi_2}^2\right)
+ {m^2\over 2}(\vec{\phi_1}^2+\vec{\phi_2}^2) +
u (\vec{\phi_1}^2+\vec{\phi_2}^2)^2 + v
(\vec{\phi}_1\wedge \vec{\phi}_2)^2}
\end{equation}
The $\beta$ functions for the couplings $u$ and $v$ are determined in a double expansion in these couplings and in $\epsilon=4-d$. The result of this analysis, at dominant order~\cite{bailin,garel,yosefin}, is that the RG equations obtained have {\it no} stable fixed point.

\subsection{A critical appraisal} 

 Frustrated magnets thus present a situation in which, contrary to the
 $O(N)$ case, the different perturbative approaches performed around the
lower and upper critical dimensions conflict when naively extrapolated to
three dimensions.  Since there is no indication that any of these
different perturbative results should fail in their respective domain of
validity -- i.e. for small $\epsilon=d-2$ and small $\epsilon=4-d$ --
this situation raises two problems. First, a practical one, which is the determination of
 the critical behaviour in three dimensions. An high-order double expansion in the coupling constants and in
$\epsilon=4-d$ or a direct perturbative computation in three dimensions,
which are both possible on the LGW model, can, {\it a priori}, deal with such
a question. This depends, however, on the ability to extract reliable
predictions out of the asymptotic series obtained, for example, by Borel
resummation. But there is no general answer to this question. In fact, it
now appears, from several recent studies, that in the case of frustrated magnets the summability of the series is weak, leading to inaccurate or ambiguous
 results~\cite{antonenko3,loison1}. A second, and more fundamental, problem is the
question of the matching between the predictions of the two field
theoretical perturbative approaches: how can the existence of a non trivial fixed point
around two dimensions be reconciled with the absence of such a fixed
point around four dimensions ?  Such a problem is, in fact, common to 
several field theories, relevant for type II superconductors, 
electroweak-phase transitions, smectic liquid cristal, He$_3$, etc. In this sense, the situation observed in the vectorial $O(N)$ model  where a non trivial fixed point exists everywhere between two and four dimensions appears to be the exception rather than the
 rule. Moreover, in the case of frustrated systems one encounters a more striking problem: there is an incompatibility between the symmetries of the NL$\sigma$ and
LGW models. Indeed, as said above, at the fixed point the symmetry of the NL$\sigma$ model action is enlarged to $O(4)$, a phenomenon that {\it cannot} occur within the 
LGW perturbative approach.  This raises serious doubts on the perturbative analysis of
the LGW model away from $d=4$. Reciprocally, the {\it perturbative}
analysis of the NL$\sigma$ model, based on a Goldstone mode expansion,
predicts an $O(4)/O(3)$ fixed point, in contradiction with the
perturbative LGW results  and the experimental and numerical situation in
$d=3$. All this suggests that  nonperturbative features could play a
major role and thus imposes to go beyond the  standard perturbative
approaches.

\section{Exact Renormalization Group Approach}

To clarify  this situation we have employed an approach based on the concept of
effective average action~\cite{wetterich2,berges3}, $\Gamma_k[\phi]$, which is a coarse
 grained free energy where only fluctuations with momenta $q\ge k$ have been integrated
 out. The field $\phi$ stands here for an average order parameter at the -- running -- scale $k$, the analog of a magnetization at this scale. In the limit $k\to 0$, all modes
 have been taken into account and $\Gamma_k$ identifies with the usual effective action $\Gamma$. At the scale $k=\Lambda$ -- the overall cut-off -- no fluctuation has been
 integrated and $\Gamma_k$ identifies with the microscopic action. The $k$-dependence of $\Gamma_k$ is controlled by an exact evolution equation~\cite{berges3,wetterich1,morris1}:
\begin{equation}
{\partial \Gamma_k\over \partial t}={1\over 2} \hbox{Tr} \left\{(\Gamma_k^{(2)}+R_k)^{-1}
 {\partial R_k\over \partial t}\right\}\ ,
\label{renorm}
\end{equation}
where $t=\ln \displaystyle {k / \Lambda}$. The trace has to be understood as a momentum
 integral as well as a summation over internal indices. In Eq.(\ref{renorm}), $\Gamma_k^{(2)}$ is  the {\it exact field-dependent} inverse propagator -- i.e. the
 second derivative of $\Gamma_k$ -- and $R_k$ is the effective infrared cut-off which suppresses the propagation of modes with momenta $q<k$.
 A convenient cut-off is provided by: $R_k(q)=Z_k
 q^2/(\exp(q^2/k^2)-1)$, where $Z_k$ is the field renormalization. The effective average action $\Gamma_k$ is a functional invariant under the symmetry group of the
system and thus includes all powers of all invariants -- and their derivatives -- built from the average order
parameter. Thus, Eq.(\ref{renorm}) is a nonlinear functional  equation,  too
difficult to be solved exactly in general. Therefore $\Gamma_k$ must be truncated. One -- fruitful --  possibility, is to perform  an
expansion of $\Gamma_k$ around the minimum in order to  keep a finite number of monomials
in the invariants  while including the most relevant derivative terms~\cite{berges3}.
We have considered the following truncation~{\cite{tissier2}:
\begin{equation}
\Gamma_k= \displaystyle \int d^dx \left\{{Z_k\over 2}  \nabla\phi_{ab}\nabla \phi_{ab}+ {
 \omega_k\over 4}\ (\epsilon_{ab}\phi_{ca} \nabla \phi_{cb})^2 +{\lambda_k\over
4}\left({\rho\over
 2}-\kappa_k\right)^2+{\mu_k\over 4}\ \tau\right\}
\label{action}
\end{equation}
where $\left\{\omega_k, \lambda_k, \kappa_k,
\mu_k,Z_k\right\}$  are the coupling constants that parametrize the  evolution of the model with
 the scale, while $\rho={\hbox{Tr}}\ ^{t}\phi\phi$ and
$\tau={1\over 2}{\hbox{Tr}} (^{t}\phi\phi)^2-{1\over
 4}({\hbox{Tr}}\ ^{t}\phi\phi)^2$  are the two independent $O(N)\otimes O(2)$ invariants built from the average order parameter. Note that, since we have considered the
generalization of the initial Heisenberg spin system to $N$-component spins, the order parameter
consists in a real $N\times 2$ matrix  $\phi_{ab}$.  Apart from the ``current term''
$(\epsilon_{ab}\phi_{ca} \nabla \phi_{cb})^2$ -- the terms in ({\ref{action}) are those appearing in
the usual
 LGW action that realizes the symmetry breaking scheme of frustrated magnets. Indeed for
 $\lambda_k$ and $\mu_k \ge 0$, the minimum of the action is realized by a configuration of
 the form $\phi_{ab}^{min}=\sqrt{\kappa_k} R_{ab}$, where $R_{ab}$ is a matrix built with
 two orthonormal $N$-component vectors. The symmetry of this minimum is a product of a
 diagonal $O(2)$ group and a residual $O(N-2)$ group. The symmetry breaking scheme is thus
 $O(N)\otimes O(2)\to O(N-2)\otimes O(2)_{diag}$. Note that for
 $\phi_{ab}=\phi_{ab}^{min}$ one has: $\rho=2\kappa_k$ and $\tau=0$ so that Eq.(\ref{action}) corresponds indeed to a quartic expansion around the minimum. At this
 minimum, the spectrum of the model, consists in $2N-3$ Goldstone modes and three massive modes: one singlet of mass $m_1=\kappa_k\lambda_k$ and  one doublet of mass $m_2=\kappa_k\mu_k$ which correspond to fluctuations of the relative angle and  of the
 norms of the two vectors $\vec{\phi}_1$ and $\vec{\phi}_2$. Let us finally stress that, without the current term, the truncation Eq.(\ref{action}) is however not sufficient in
 our case. This term plays a crucial role since, for $N=3$, it allows the model to enlarge
 its symmetry from $O(3)\otimes O(2)$ to $O(3)\otimes O(3)\sim O(4)$ at the fixed point,
 above $d=2$. The current term is systematically discarded in the perturbative treatment
 of the LGW model around four dimensions, for the -- correct -- reason that it is power-counting irrelevant. Here we can include it in our {\it ansatz} since it is anyway
 present in the full effective action $\Gamma_k$. In fact, we {\it must} include it since
 it becomes relevant somewhere between two and four dimensions, the formalism we use being in charge to decide where it is important.  

 We have derived the flow equations for the different coupling constants $\kappa_k$, $\lambda_k$, $\mu_k$, $\omega_k$ and  $Z_k$ using Eq.(\ref{renorm}) and Eq.(\ref{action})~\cite{tissier2}. The explicit recursion equations for the dimensionless renormalized coupling constants $g_i^r$ are too lengthy to be displayed here and not particularly illuminating~\cite{site}. Their general structure is given by:
\begin{equation}
{dg_i^r\over dt}=D g_i^r + F_i\left[\{g_j^r\}, \{T_j[m_j^r]\}\right]
\label{rgeq}
\end{equation}
where $D$ stand for the dimension -- including the anomalous dimension -- of the coupling $g_i^r$. In Eq.(\ref{rgeq}), the functions $F_i$ involve the whole set of coupling 
 constants $\{g_j^r\}$. More important for our purpose is that the flow equations also
 involve the whole spectrum -- Goldstone as well as massive excitations -- of the theory.
 These masses occur as the arguments of some ``threshold''  functions $\{T_j\}$ that
 govern the coupling or decoupling of the massive degrees of freedom. This is precisely
 this aspect that allows the method to go beyond the pure Goldstone mode expansion
performed with the NL$\sigma$ model around $d=2$. 

In practice, we have studied the fixed point structure of the flow equations (\ref{rgeq})
 everywhere between two and four dimensions. Immediately around two dimensions, the fixed
 point masses $m_1^{*r}$ and $m_2^{*r}$ diverge as $1/\epsilon$. This means a decoupling
 of the excitations associated to the fluctuations of the modulus and of the relative
 angle of $\vec{\phi}_1$ and $\vec{\phi}_2$. The physics is, as expected, fully controlled
 by the Goldstone modes. The flow equations (\ref{rgeq}) then degenerate to those of the
 $O(N)\otimes O(2)\to O(N-2)\otimes O(2)_{diag}$ NL$\sigma$ model at one loop. For $N=3$,
 they admit a fixed point for which the model is
 $O(4)$ symmetric, attesting the validity of the perturbative NL$\sigma$ model approach.
 However, the flow equations actually admit another -- unstable -- fixed point, absent in
 the NL$\sigma$ model approach, which moves toward the stable one while $d$ is increased. 
 At a critical dimension $d_c\simeq 2.87$, the two fixed points collapse and disappear.
 Above $d_c$ no other stable fixed point has been found, in agreement with the LGW
 approach around four dimensions. Note also that we have checked that, around four
 dimensions, an expansion of our equations in powers of the coupling constants allows to
 recover exactly the one-loop RG equations found within the LGW approach. The transition
 in $d=3$ is thus, rigorously speaking, first order.  However, the proximity of $d_c$ with
 $d=3$ lets open the possibility  of a very weak first order phase transition in this
 dimension. We have checked that there indeed exists, in $d=3$, a range of parameters for
 which the RG flow is very slow. As a consequence the correlation length, although finite,
 is very large and a pseudo-scaling behaviour -- an ``almost second order phase transition
 ''~\cite{zumbach7} -- can be observed. Note however that $\xi$ remaining always finite, a
 true universality should not be expected. This could explain both the observation of a
 scaling behaviour as well as the existence of a broad  spectrum of effective critical
 exponents in experiments and in numerical simulations. To test this scenario, we have
 computed effective exponents by linearizing the flow equations around the minimum, taken
 into account the $\phi^6$-like terms in our {\it ansatz}. We have checked that the
 inclusion of monomials of order six in the fields is sufficient  to ensure the stability
 of our results. We have found: $\nu=0.53$, $\gamma=1.03$ and $\beta=0.28$, which lie in
 between the various sets of exponents found experimentally and numerically. At this
 stage, one has to mention a work of Pelissetto {\it et~al.} who have studied the
 $O(N)\otimes O(2)\to O(N-2)\otimes O(2)_{diag}$ LGW model at six-loop order in fixed
 dimension $d=3$ and have reached a somewhat different conclusion~\cite{pelissetto1}. From
 their analysis, they have concluded that there exists a non trivial fixed point in the
 Heisenberg case. This is in contrast with a previous three-loop calculation in which no
 fixed   point was found~\cite{antonenko3,loison1}. The question is thus that of the
 stability of the perturbative results with respect to the number of loops or to the Borel
 resummation procedures.  However, amazingly, the exponents found by Pelissetto {\it et~al.} -- $\nu=0.55(3)$, $\gamma=1.06(5)$ and  $\beta=0.30(2)$ -- are very close to our
 pseudo-exponents, making the question of the existence of a true fixed point somewhat
 academic. Finally, we have found a true stable fixed point in $d=3$ for $N$ larger than a
 critical value $N_c(d=3)\simeq 5$. For $N=6$, using $\phi^{10}$-like terms, we get
 $\nu=0.707$ and $\gamma=1.377$ that lie in between the error bars of the Monte Carlo
 data~\cite{loison1} which display a clear second order behaviour with  $\nu=0.700(11)$
 and $\gamma=1.383(36)$. Note that no result has been reported within the six-loop
 perturbative calculation in this case.

Using an exact renormalization group approach, we have reached a global understanding of the physics of frustrated Heisenberg magnets including a matching between the previous perturbative predictions performed around the upper and lower critical dimensions and a good agreement with experimental and numerical data for the critical physics in $d=3$.

\end{document}